\documentclass[prl,aps,twocolumn,superscriptaddress]{revtex4-1}
\usepackage{graphicx,color}
\usepackage{amsthm}
\usepackage{amsfonts}
\usepackage{algorithmic}
\usepackage{enumerate}
\usepackage{latexsym}
\usepackage{amsmath}
\usepackage{amssymb}
\usepackage[colorlinks=true,citecolor=blue,linkcolor=blue]{hyperref}

\usepackage{dcolumn}
\usepackage{bm}

\usepackage[T1]{fontenc}
\usepackage{mathptmx}

\begin{document}
\title{Magnetic phase transition in disordered interacting Dirac fermion systems via the Zeeman field}

\author{Jingyao Meng}
\affiliation{Department of Physics, Beijing Normal University, Beijing 100875, China}
\author{Lufeng Zhang}
\affiliation{School of Science, Beijing University of Posts and Telecommunications, Beijing 100876, China\\}
\author{Tianxing Ma}
\email{txma@bnu.edu.cn}
\affiliation{Department of Physics, Beijing Normal University, Beijing 100875, China}
\affiliation{Beijing Computational Science Research Center, Beijing 100193, China}
\author{Hai-Qing Lin}
\affiliation{Beijing Computational Science Research Center, Beijing 100193, China}
\affiliation{Department of Physics, Beijing Normal University, Beijing 100875, China}

\begin{abstract}
Using the determinant quantum Monte Carlo method, we investigate the antiferromagnetic phase transition that is induced by the Zeeman field in a disordered interacting two-dimensional Dirac fermion system. At a fixed interaction strength $U$, the antiferromagnetic correlation is enhanced as the magnetic filed increases, and when the magnetic field is larger than a $B_{c}(U)$, the antiferromagnetic correlation shall be suppressed by the increased magnetic field.
The impact of Zeeman field $B$, Coulomb repulsion $U$ and disorder $\Delta$ is not isolated.
The intensity of magnetic field effect on the antiferromagnetic correlation shall be strongly suppressed by disorder.
Differently, it will be promoted by weak interaction, but when $U$ becomes larger than $U_{c}=4.5$, the increased interaction will suppress the intensity of this effect, and here $U_{c}=4.5$ coincides with the critical strength inducing the metal-Mott insulator transition in clean system.
Moreover, at a fixed magnetic field $B$, strong interaction shall suppress the antiferromagnetic phase rather than promote it.
\end{abstract}

\pacs{71.10.Fd, 74.72.-h, 67.85.-d, 05.10.-a}

\maketitle

\section{Introduction}
According to studies on the widely investigated Dirac fermion systems, graphene is one of the most promising 2D materials\cite{GUPTA201544,Nat.Mater.6.183} due to its unique characteristics, such as excellent electrochemical performance\cite{LI2018845} and ultrahigh electrical conductivity\cite{https://doi.org/10.1002/adma.200800757,PhysRevB.83.165113}. With the emergence of a series of novel phenomena, such as the Pomeranchuk effect\cite{rozen2021entropic}, tunable strongly coupled superconductivity\cite{park2021tunable} and the spin-Hall effect\cite{PhysRevLett.107.096601}, the honeycomb lattice that is applied in strong correlation physics is expected to be associated with breakthrough results.
To increase the similarity between the model and the actual material and enhance the reliability of the results, we considered both interaction and disorder in our research. Their interplay led to more complex physical mechanisms, thereby inducing a series of valuable discoveries. For example, the hopping disorder closes the Mott gap\cite{PhysRevB.58.15314} and induces a novel nonmagnetic insulating phase which is found to emerge from the zero-temperature quantum critical point\cite{PhysRevLett.120.116601}, and disordered impurities can drive the ordered state by electron-mediated interaction at a transition temperature\cite{doi:10.1021/acs.nanolett.1c02714}. For this reason, the study of disordered interacting Dirac fermion systems, with the graphene lattice considered as an example, has far-reaching significance and value.

As the basic property of crystal lattices, magnetism has always received extensive attention due to its reflection on the physics of systems. For example, transport phase transitions are accompanied by the antiferromagnetism \cite{PhysRevLett.120.116601,PhysRevLett.93.016406}, or the magnetostriction is induced by the broken time reversal symmetry\cite{Rotter_2006}. Since a variety of fascinating properties are closely connected with magnetism (such as superconductivity\cite{PhysRevLett.120.247002,PhysRevB.82.014521} and topological band insulator properties\cite{PhysRevB.84.205121}), research on magnetism has become increasingly richer.
As a system that has particle-hole symmetry at half-filling, the honeycomb lattice provides an excellent research platform for promoting interesting phenomena in this dynamic field\cite{PhysRevLett.126.157201,PhysRevLett.126.107205}, especially antiferromagnetic (AFM) phase transitions, which have long been a central and controversial issue. The Zeeman field can create a better environment for studying this topic: it will couple only to the spin, not to the orbital motion of electrons\cite{PhysRevLett.90.246401,PhysRevB.80.115428}, and polarize the graphene carriers to affect the density of states\cite{PhysRevB.80.075417}. In the continuum limit, an in-plane magnetic field has been proven to facilitate spontaneous symmetry breaking\cite{kharzeev2006spin}. Therefore, how the magnetic field, interaction and disorder affect one another and work together is an essential and interesting issue.

In this paper, we use the exact determinant quantum Monte Carlo (DQMC) method and study the Hubbard model on a honeycomb lattice. By examining the staggered transverse AFM structure factor, we focus on the AFM phase in a more interesting direction parallel to the lattice plane. The Zeeman field initially induces and subsequently suppresses the AFM phase in this direction, which is different from the strong inhibitory effect in the vertical direction. The interplay of interaction, disorder and the parallel field is reflected as complex magnetic effects: First, the effect of disorder is confined to a limited range of field strengths. Second, the Zeeman field causes an effect that is contrary to conventional understanding, namely, the AFM phase will be suppressed by the strong interaction.
These impacts are not unidirectional. Increasing disorder renders the influence of the parallel field more hidden, and sufficiently strong disorder leads to the absence of the AFM phase regardless of the magnetic field. With increasing interaction, the induction-inhibition effect of the magnetic field on the staggered transverse AFM structure factor becomes more likely to be produced, and the effect intensity also changes -- it initially increases and subsequently decreases. Notably, the interaction strength at which the magnetic field effect is most obvious is the critical value that induces the AFM Mott insulating phase transition in the clean system. Overall, complex and interesting magnetism occurs in the disordered interacting Dirac Fermi system via a magnetic field, which is induced by the combined effect of three factors.

\section{Model and method}
The Hamiltonian of the disordered Hubbard model on a honeycomb lattice in the presence of a magnetic field is defined as follows:
\begin{eqnarray}
\label{Hamiltonian}
\hat{H}&=&-\sum_{\langle{\bf ij}\rangle\sigma}t_{\bf ij}(\hat{c}_{{\bf i}\sigma}^\dagger\hat{c}_{{\bf j}\sigma}
+\hat{c}_{{\bf j}\sigma}^\dagger\hat{c}_{{\bf i}\sigma})+U\sum_{\bf j}(\hat{n}_{{\bf j}\uparrow}-\frac{1}{2})(\hat{n}_{{\bf j}\downarrow}-\frac{1}{2}) \nonumber\\
&&-\sum_{{\bf j}\sigma}(\mu-{\sigma}B)\hat{n}_{{\bf j}\sigma},
\end{eqnarray}
where $\hat{c}_{{\bf j}\sigma}^\dagger(\hat{c}_{{\bf j}\sigma})$ is the spin-$\sigma$ electron creation (annihilation) operator at site $\bf i$ and $\hat{n}_{{\bf i}\sigma}=\hat{c}_{{\bf i}\sigma}^\dagger\hat{c}_{{\bf i}\sigma}$ is the occupation number operator. Here, $t_{\bf ij}$ is the \textit{nearest-neighbor} (NN) hopping integral, $U > 0$ is the onsite Coulomb repulsive interaction, $\mu$ is the chemical potential, and $B$ is the Zeeman field along the $x$ direction which is  parallel to the lattice plane (thus, orbital contributions are not generated~\cite{PhysRevLett.90.246401}). By selecting the hopping parameters $t_{\bf ij}$ from the probability distribution ${\cal P}(t_{\bf ij})$ = 1$/\Delta$ for $t_{\bf ij} \in [t - \Delta/2,t + \Delta/2]$ and setting them to zero otherwise, disorder is introduced into the system. $\Delta$ describes the strength of the disorder, and $t = 1$ sets the energy scale in the following. By choosing $\mu=0$, we obtained a half-filled system with particle-hole symmetry~\cite{PhysRevLett.83.4610}.

We adopt the DQMC method~\cite{PhysRevB.40.506} to study the magnetic phase transition in the model that is defined by Eq.~\eqref{Hamiltonian}, in which the Hamiltonian is mapped onto free fermions in 2D+1 dimensions that are coupled to space- and imaginary-time-dependent bosonic (Ising-like) fields. By using Monte Carlo sampling, we can carry out the integration over a relevant sample of field configurations, which are selected until the statistical errors become negligible.
The discretization mesh $\Delta\tau$ of the inverse temperature $\beta = 1/T$
should be small enough to ensure that the qualified Trotter errors are less than those that are associated with statistical sampling.
This approach enables us to compute static and dynamic observables at a specified temperature $T$. Due to the particle-hole symmetry even in the presence of the hopping-quenched disorder, the system avoids the infamous minus-sign problem, and the simulation can be performed at a large enough value of $\beta$ to obtain properties that converge to the ground-state properties~\cite{PhysRevLett.120.116601,PhysRevLett.115.240402}. We choose an $L = 12$ honeycomb lattice with periodic boundary conditions, for which the total number of sites is $N =2\times$$L^2$. In the presence of disorder, we average over 20 disorder realizations~\cite{PhysRevLett.120.116601,PhysRevB.54.R3756,PhysRevB.76.144513,PhysRevB.47.7995,IdealFermiSystems}.

\begin{figure}[t]
\includegraphics[scale=0.5]{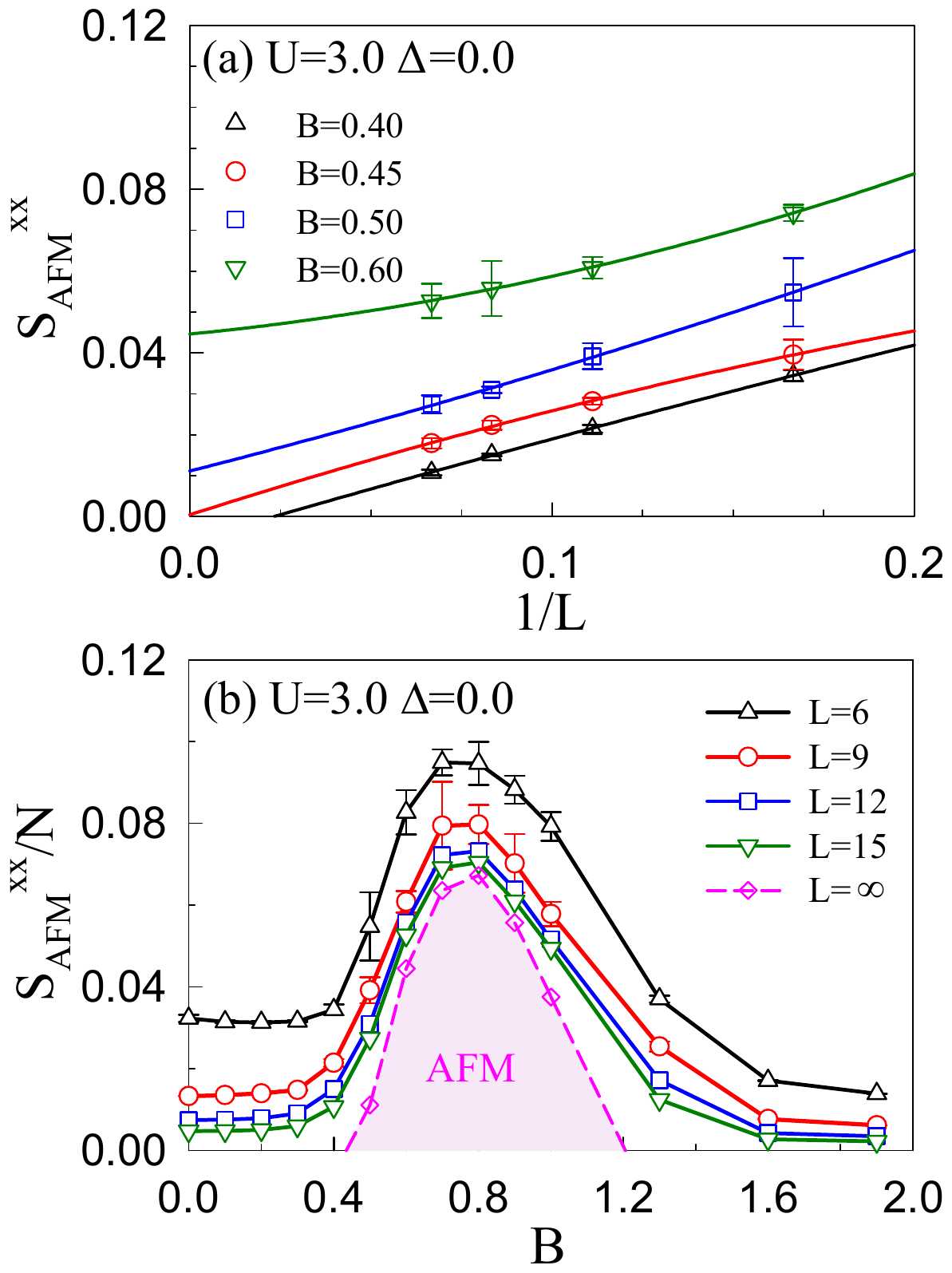}
\caption{\label{FigSxB} Staggered transverse antiferromagnetic (AFM) structure factor $S_{\rm AFM}^{\rm xx}$: (a) As a function of the lattice size $L$ for various values of the magnetic field $B$. As the magnetic field increases, $S_{\rm AFM}^{\rm xx}$ is increased at each $L$. As the curve intercept gradually increases from zero to positive, the system reaches the AFM phase. The critical strength for the magnetic phase transition is approximately $B = 0.45$.
(b) As a function of $B$ for various values of $L$. The pink curve, which represents $L = \infty$, namely, $1/L = 0$, is made up of intercepts that were obtained by $S_{\rm AFM}^{\rm xx}(L)$ curve fitting under each magnetic field. This curve shows the the appearance/disappearance of the AFM phase. The induction of the AFM phase requires a large value of $B$ and will be eliminated by the magnetic field if its strength continues to increase. Calculations are performed on $2\times$$L\times$$L$ lattices for $U = 3.0$ and $\Delta = 0.0$.}
\end{figure}

To study the magnetic phase transition, particularly to characterize the AFM phase, we compute the staggered transverse antiferromagnetic structure factor in the direction parallel to the lattice plane as
\begin{equation}
 S_{\rm AFM}^{\rm xx} = \frac{1}{N}\sum_{i,j}(-1)^{(i+j)} \left(S_i^{x}S_j^{x} + S_i^{y}S_j^{y}\right),
\end{equation}
where $S_i^{x}$ ($S_i^{y}$) is the $x$ ($y$)-component spin operator and the phase factor is $+1$($-1$) for sites $i$,$j$ that belong to the same (different) sublattices of the honeycomb structure. Similarly, the longitudinal structure factor $S_{\rm AFM}^{zz} \equiv (1/N)\sum_{i,j}(-1)^{(i+j)} S_i^{z}S_j^{z}$ describes the magnetic order in the $z$ direction. Finally, we introduce the parameter $P=|n_{\downarrow}-n_{\uparrow}|/(n_{\downarrow}+n_{\uparrow})$ to study the spin polarization of electrons, where $n_{\downarrow}$ and $n_{\uparrow}$ are the averaged spin-resolved densities of the corresponding number operators in Eq.~\eqref{Hamiltonian}, and $\uparrow$ ($\downarrow$) is parallel (antiparallel) to the direction of $x$ or $z$ direction. In addition, since the AFM phase disappears at high temperatures, we choose $T = 1/12$, which is small enough to avoid this temperature effect, as shown in Fig.~\ref{FigPB}(c).

\section{Results and Discussion}

\begin{figure*}[t]
\includegraphics[scale=0.5]{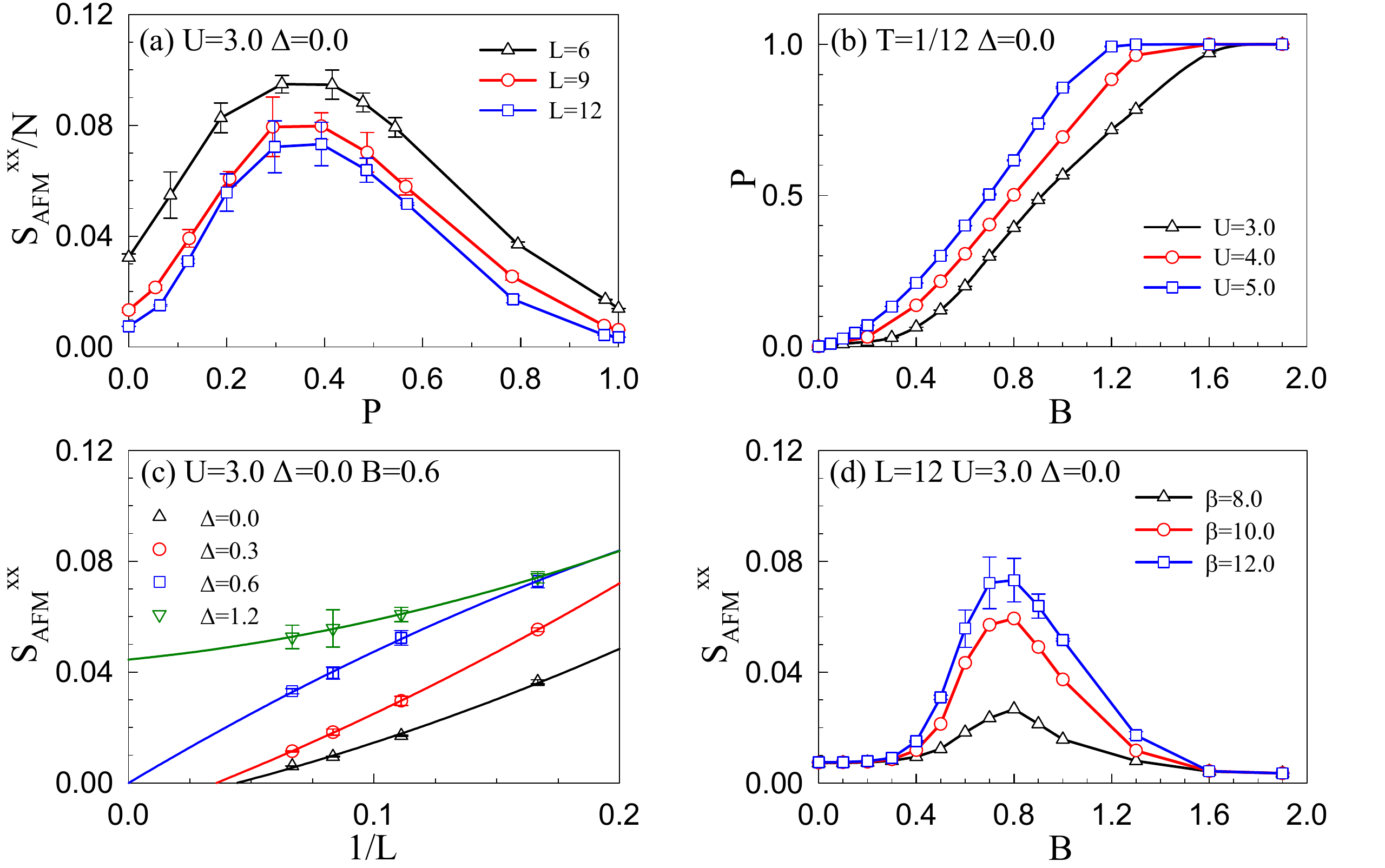}
\caption{\label{FigPB} (a) The staggered transverse AFM structure factor $S_{\rm AFM}^{\rm xx}$ as a function of the spin polarization $P$ at various values of $L$. With $P$ as the abscissa, the curve has a more dramatic change.
(b) $P$ as a function of $B$, where the function relationship depends on the interaction $U$: the $P(B)$ curve under a large value of $U$ enters the ascending stage faster and reaches 1 (the fully spin-polarized state) first.
(c) $S_{\rm AFM}^{\rm xx}$ as a function of $L$ for various values of the inverse temperature $\beta$. The temperature effect is exhibited as the AFM phase is eliminated by the large $T$, which is accompanied by the disappearance of the intercept.
(d) $S_{\rm AFM}^{\rm xx}$ as a function of $B$ for various values of $\beta$. The influence of $T$ is obvious as $S_{\rm AFM}^{\rm xx}$ increases with increasing $\beta$ under each $B$.}
\end{figure*}

\begin{figure*}[t]
\includegraphics[scale=0.5]{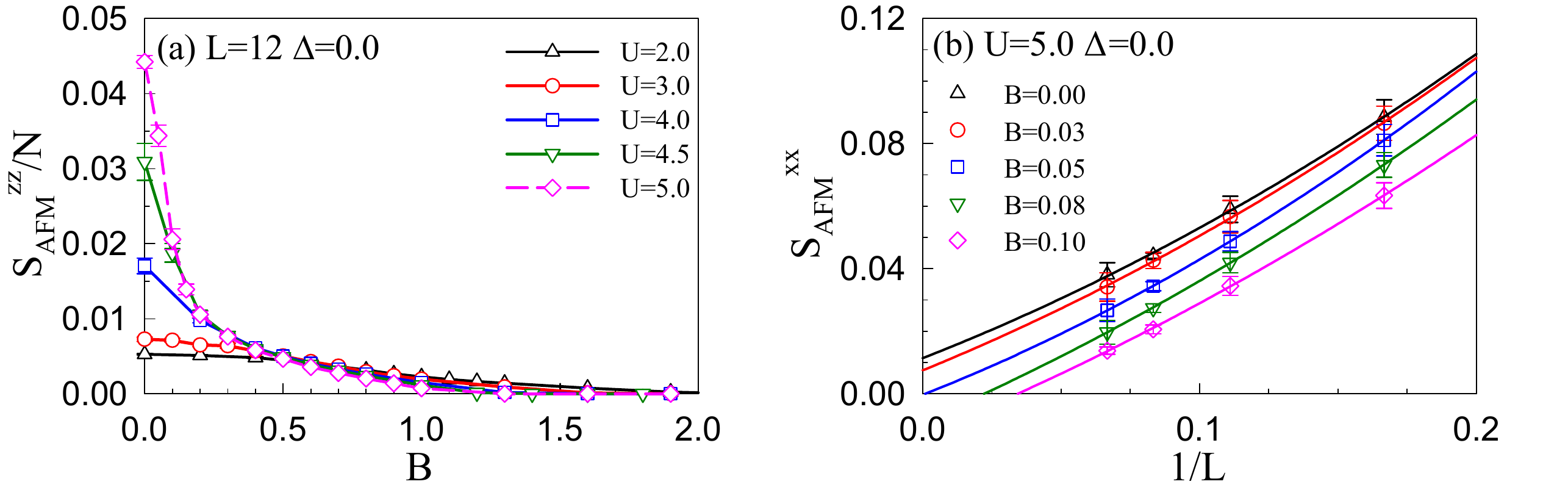}
\caption{\label{FigSzB} Staggered longitudinal antiferromagnetic structure factor $S_{\rm AFM}^{\rm zz}$ in the clean system: (a) As a function of $B$ at various values of $U$. $S_{\rm AFM}^{\rm zz}$ are effectively suppressed and reduced to 0 by the magnetic field.
(b) As a function of $L$ for various values of $B$. The suppression of $B$ in the AFM phase is more clearly shown through the intercept of the $S_{\rm AFM}^{\rm zz}(L)$ curve, which quickly drops to 0 even under a weak magnetic field.}
\end{figure*}

We start from a clean system without disorder, where a parallel magnetic field induces the antiferromagnetic (AFM) transition in the system.
The lattice is insulating at interaction $U/t = 3$, disorder $\Delta = 0$ and magnetic field $B = 0.4$ (see Appendix~\ref{app:DCcon}). At this time, as shown in Fig.~\ref{FigSxB}(a), the staggered transverse AFM structure factor $S_{\rm AFM}^{\rm xx}$ tends to 0 when $1/L \rightarrow 0$ (namely, $L \rightarrow \infty$), which suggests that the structure factor is not extensive, thereby resulting in only short-range ordering. When $B$ grows to a sufficiently large value, such as $B = 0.6$, the $S_{\rm AFM}^{\rm xx}$ curve is predicted to have a positive intercept at $1/L = 0$, which corresponds to the appearance of the AFM phase.
Interestingly, the induction of AFM by the magnetic field does not happen immediately, even though $B$ is large enough to induce the
MIT (more details are provided in the Appendix~\ref{app:DCcon}). In addition, in Fig.~\ref{FigSxB}(b), the pink curve is made up of intercepts of $S_{\rm AFM}^{\rm xx}(L)$ under several values of $B$; hence, the AFM phase only exists in the area where the curve is above the horizontal axis. The magnetic field hardly affects the value of $S_{\rm AFM}^{\rm xx}$ until $B = 0.4$. This indicates that a fairly strong magnetic field is needed to cause symmetry breaking of the lattice. Notably, $S_{\rm AFM}^{\rm xx}$ is not always positively correlated with the magnetic field: With an increase in large $B$, the system becomes increasingly close to full-spin polarization, thus $S_{\rm AFM}^{\rm xx}$ continues to decrease until $= 0$.

\begin{figure*}[t]
\includegraphics[scale=0.5]{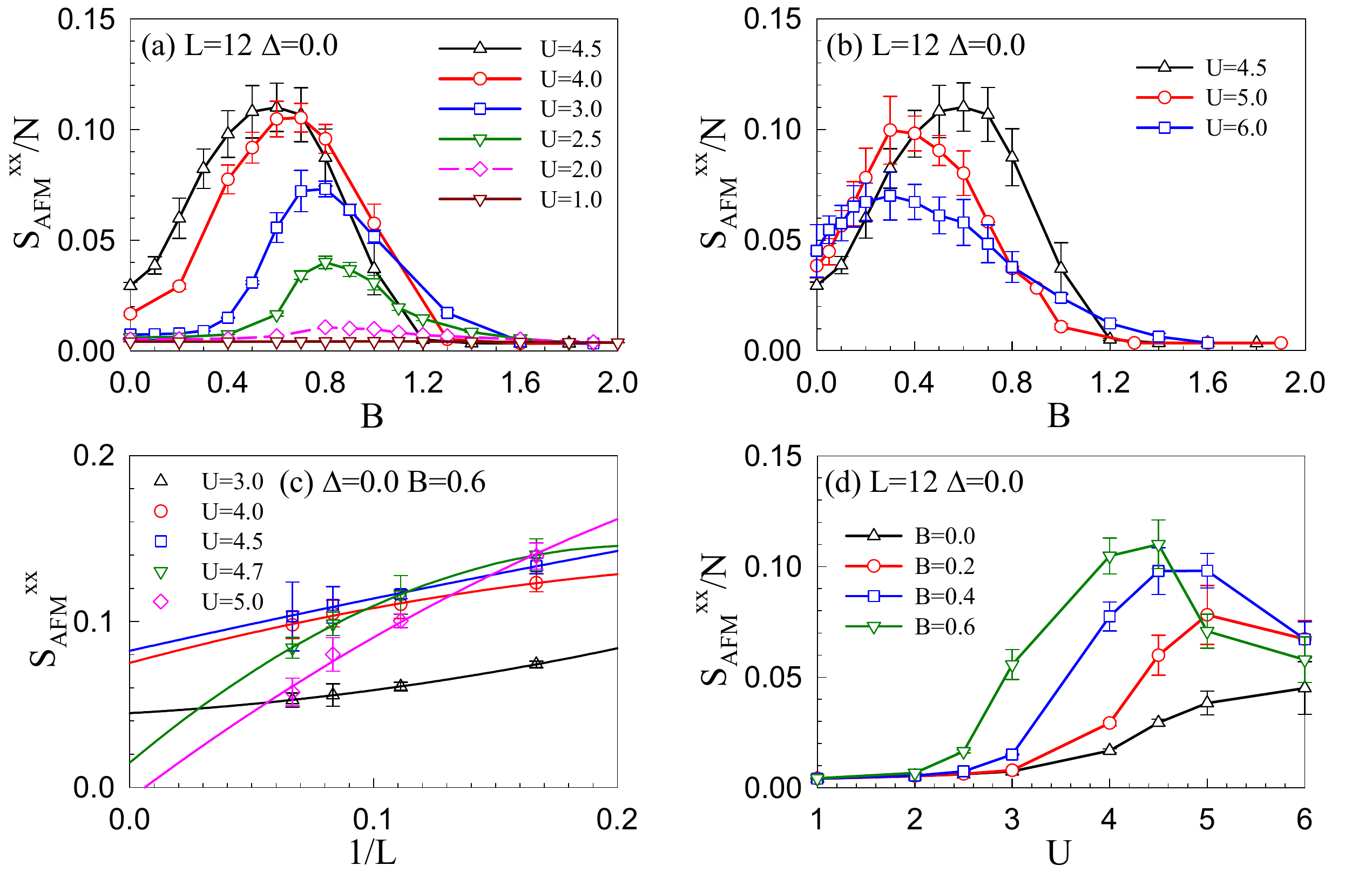}
\caption{\label{FigSxU} $S_{\rm AFM}^{\rm xx}$ as a function of $B$ at $L = 12$, $\Delta = 0.0$, and $U$ in the following ranges: (a) $U = 1.0$ to $4.5$. Each $S_{\rm AFM}^{\rm xx}(B)$ curve has a special magnetic field strength $B_{c}$ for the peak value $S_{max}$. As $U$ increases from 0 to 4.5, $S_{max}$ always increases, which is accompanied by a decrease in $B_{c}$.
(b) $U =$ 4.5 to 6.0. When the increased interaction exceeds 4.5, although $B_{c}$ continues to decrease, $S_{max}$ has a decreasing trend, in contrast to the previous stage. The point that separates the two stages is at $U = 4.5$, which is the critical value for the interaction-induced AFM phase transition.
(c) $S_{\rm AFM}^{\rm xx}$ as a function of $L$ at $\Delta = 0.0$. At $B = 0.6$, starting from the AFM phase at $U = 3.0$, the effect of the interaction on the AFM phase presents a transition from prompting to inhibiting (the intercept of the $S_{\rm AFM}^{\rm xx}$ curve initially increases and subsequently decreases).
(d) $S_{\rm AFM}^{\rm xx}$ as a function of $U$ at various values of $B$. As $B$ increases, the effect of the interaction promoting the AFM phase is gradually reversed. When $B \neq 0$, the $S_{\rm AFM}^{\rm xx}(U)$ curve shows a trend of rising initially and subsequently falling instead of a monotonic increase.}
\end{figure*}

In Fig.~\ref{FigPB}(a) and (b), we show the results for $S_{\rm AFM}^{\rm xx}$ as a function of the degree of spin polarization $P$ and $P$ as a function of $B$, which accord with the results of the previous study and support our conclusions.
Because the semimetallic state under an in-plane magnetic field is considered unstable, $S_{\rm AFM}^{\rm xx}$ curves quickly enter the upward phase with $P$ as the abscissa\cite{PhysRevB.80.045412}. The phenomena in Fig.~\ref{FigSxB}(a) and Fig.~\ref{FigPB}(a) differ because the relationship between polarization and the magnetic field is not linear: as shown in Fig.~\ref{FigPB}(b), with an increase in $B$, $P$ first increases slowly, then rises gradually at a faster rate, and finally converges to 1 to reach a fully spin-polarized state. The behavior of $P$ is also affected by $U$, which is also reflected in Ref. \cite{PhysRevB.80.045412}. Notably, the increase in spin polarization by the interaction only occurs in the presence of a magnetic field. In addition, the temperature effect is shown in panels (c) and (d), where $S_{\rm AFM}^{\rm xx}$ is a function of $L$ or $B$, and the AFM phase is shown to be eliminated at high temperature.

In contrast to the interesting effect of $S_{\rm AFM}^{\rm xx}$, the parallel magnetic field is proven to effectively suppress the AFM phase in the $z$ direction perpendicular to the lattice plane. In Fig.~\ref{FigSzB}(a), we report the staggered longitudinal AFM structure factor $S_{\rm AFM}^{\rm zz}(B)$ computed across several representative interactions $U$. As $U$ increases, the transition from the (semi)metallic phase to the Mott insulating phase occurs on the graphene lattice (see Appendix). Because $U$ significantly increases the value of $S_{\rm AFM}^{\rm zz}$ at $B = 0$ in Fig.~\ref{FigSzB}(a), its ability to induce antiferromagnetism is proven, and the critical value is approximately 4.5\cite{1992U4.5,PhysRevB.72.085123,PhysRevB.101.245161}. In the direction perpendicular to the lattice plane, the spin is extremely sensitive to the parallel field, and the original magnetic order is soon destroyed by the Zeeman field, which corresponds to a rapid decline of the $S_{\rm AFM}^{\rm zz}$ curve with the introduction of $B$.
A more evident display of the magnetic field is obtained in Fig.~\ref{FigSzB}(b), where $S_{\rm AFM}^{\rm zz}$ at various values of the magnetic field $B$ is plotted as a function of $1/L$. To make the effect of $B$ more obvious, we choose the system in the AFM phase at $U = 5.0$. In contrast to $S_{\rm AFM}^{\rm xx}$, $S_{\rm AFM}^{\rm zz}$ is extremely sensitive to changes in the magnetic field: At $B = 0$, the $S_{\rm AFM}^{\rm xx}$ curve has a positive intercept on the vertical axis, and a very small magnetic field (for example, $B = 0.05$) can make the intercept tend to zero. The suppression of $S_{\rm AFM}^{\rm xx}$ by the magnetic field will continue until the value tends to zero.

\begin{figure*}[t]
\includegraphics[scale=0.5]{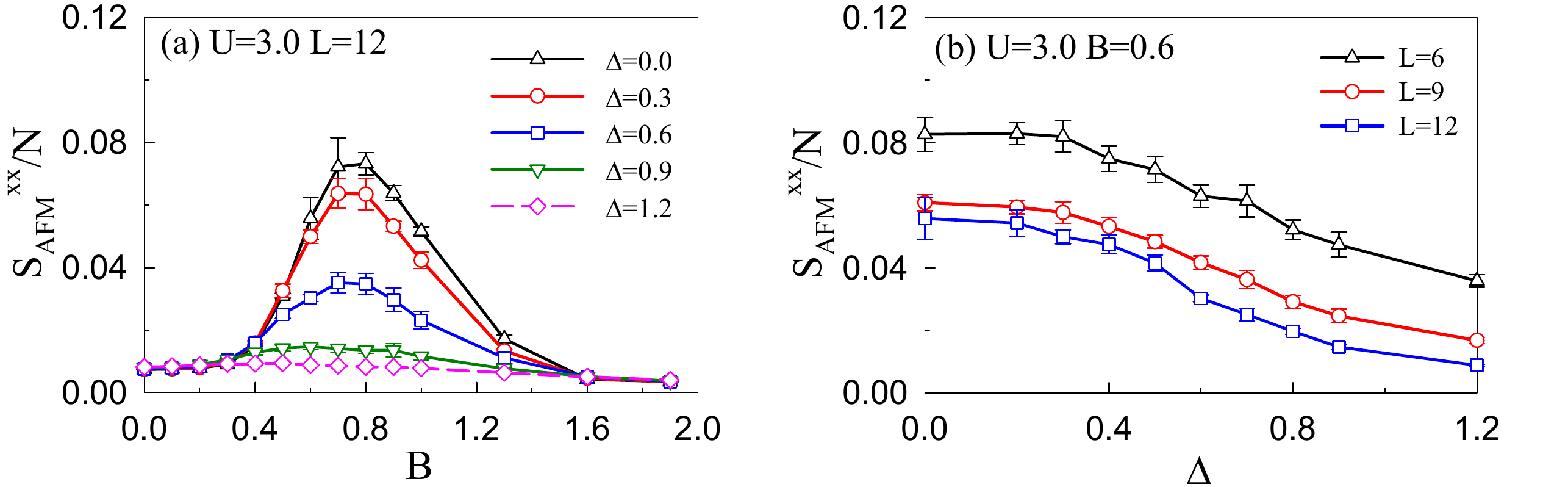}
\caption{\label{FigSxD} (a) $S_{\rm AFM}^{\rm xx}$ as a function of $B$ for various values of $\Delta$ at $U = 3.0$ and $L = 12$. As $\Delta$ increases from 0, $S_{\rm AFM}^{\rm xx}$ is suppressed, and the peak of the curve gradually disappears. When $\Delta = 1.2$, the magnetic field has almost no effect on $S_{\rm AFM}^{\rm xx}$.
(b) $S_{\rm AFM}^{\rm xx}$ as a function of $\Delta$ for various values of $L$. Starting from the AFM phase at $U = 3.0$, $\Delta = 0.0$ and $B = 0.6$ in Fig.~\ref{FigSxB}(a), an increase in $\Delta$ leads to a decrease in $S_{\rm AFM}^{\rm xx}$ under each $L$, namely, disorder fully suppresses the AFM phase.}
\end{figure*}

\begin{figure*}[t]
\includegraphics[scale=0.5]{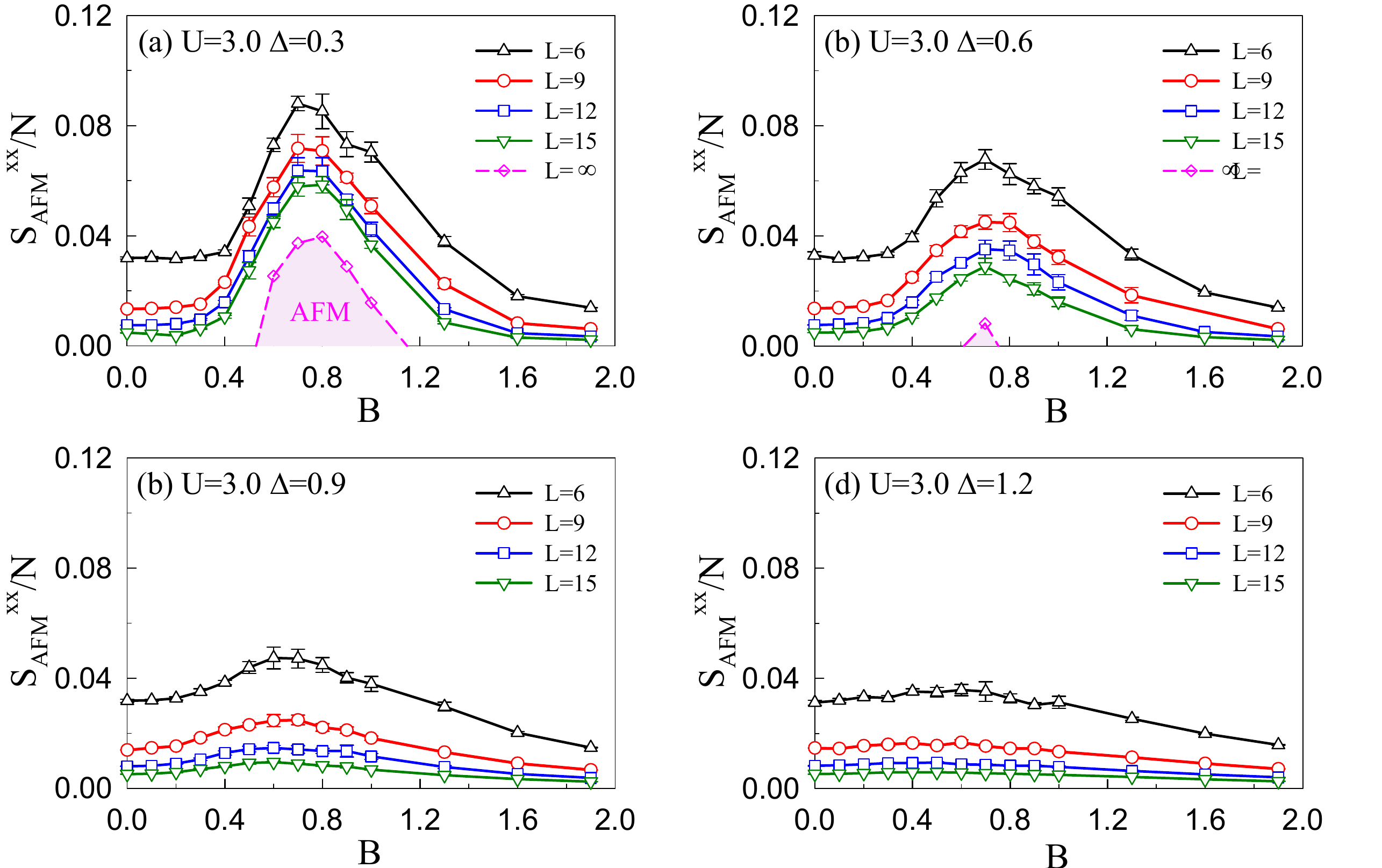}
\caption{\label{FigSxBD} $S_{\rm AFM}^{\rm xx}$ as a function of $B$ for various values of $L$. The pink curve, which represents $L = \infty$, is the combination of the values to which $S_{\rm AFM}^{\rm xx}-1/L$ curves (under each values of $B$) tend when $1/L = 0$. Therefore, its decline due to increasing $\Delta$ intuitively represents the suppression of the AFM phase by disorder. Calculations are performed at $U = 3.0$ and $\Delta =$ 0.3(a), 0.6 (b), 0.9 (c) and 1.2 (d).}
\end{figure*}

Since the suppression of $B$ on $S_{\rm AFM}^{\rm zz}$ is so obvious and direct, we focus on the more interesting transverse direction.
For antiferromagnetism, while the conflict between disorder and interaction has received widespread attention, the effect of their interplay with a magnetic field remains to be elucidated. We show the complex changes in magnetic order under the combination of interaction $U$ and a magnetic field $B$ in Fig.~\ref{FigSxU}, which shows that they substantially influence each other's effect on $S_{\rm AFM}^{\rm xx}$.

First, the interaction $U$ comprehensively affects the induction and inhibition of the AFM phase that is induced by the Zeeman field $B$, including the strength of the magnetic field effect and the effective range of $B$. As discussed previously, the magnetic field needs to be strong enough to increase $S_{\rm AFM}^{\rm xx}$, and a peak value appears in the process of induction and subsequent inhibition of AFM phase by magnetic field. We denote the magnetic field strength at this time as $B_{c}$ and the peak value as $S_{max}$. As shown in Fig.~\ref{FigSxU}(a) and (b), as $U$ increases, the value of $dS_{\rm AFM}^{\rm xx}/dB$ at $B = 0$ increases from zero, and $B_{c}$ gradually decreases. This phenomenon indicates that interaction makes it easier for the magnetic field to impact and accelerate the emergence of peaks due to its promotion on $S_{\rm AFM}^{\rm xx}$, namely, the AFM phase. In addition, $S_{max}$ is another factor that merits discussion, which is decided by not $B$ but $U$. In panels (a) and (b), we plot the intervals where $S_{max}$ increases and decreases with $U$, and their dividing point is approximately $U_{c} = 4.5$, which is proven to be the critical value for the $U-$induced AFM phase transition in the graphene system\cite{1992U4.5,PhysRevB.72.085123,PhysRevB.101.245161}.
Under weak interactions, such as the curve of $U = 1$ in panel (a), the magnetic field hardly promotes $S_{\rm AFM}^{\rm xx}$, nor can it induce an AFM phase transition. As $U$ increases, the effect of $B$ gradually becomes obvious, and $S_{max}$ increases. When $U$ reaches 4.5, the critical value, namely, $S_{max}$, reaches the maximum and gradually weakens as $U$ increases. These results all show that the interaction is closely related to the intensity of the magnetic field effect. If we use $S_{max}$ to characterize the intensity of magnetic field effect on the AFM correlations, one can see that this effect will be promoted by weak interaction but suppressed by strong interaction, where $U_{c}=4.5$ distinguishes these two interaction regions.

\begin{figure*}[t]
\includegraphics[scale=0.5]{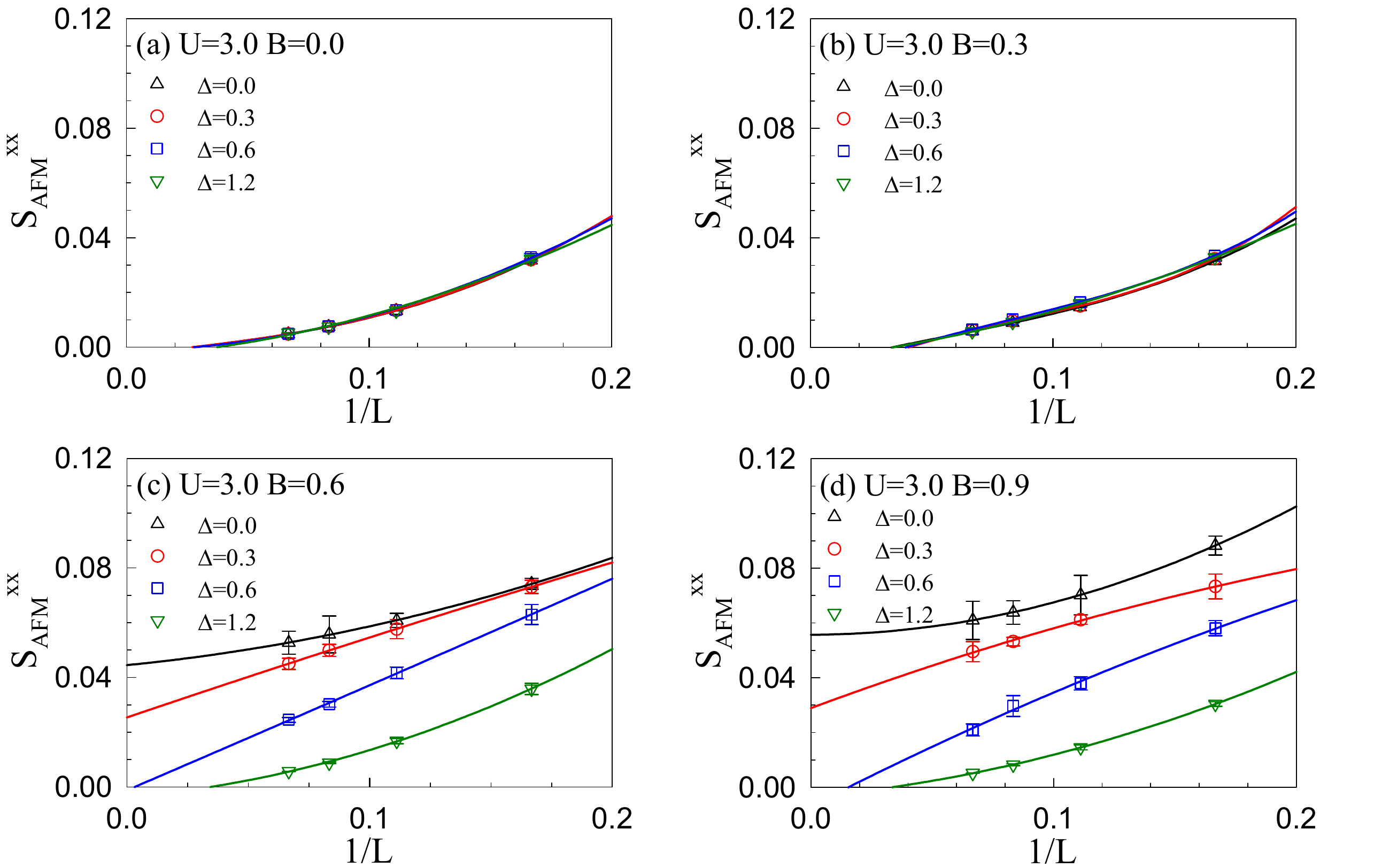}
\caption{\label{FigSxL} $S_{\rm AFM}^{\rm xx}$ as a function of $L$ for various values of $\Delta$ at $U = 3.0$. The magnetic fields in panels (a), (b), (c) and (d) are 0.0, 0.3, 0.6 and 0.9, respectively, and the systems without $\Delta$ are the metallic phase (a), band insulating phase (b) and AFM phase (c) (d), respectively. The disorder effect only appears under a strong enough $B$, which is manifested as a decrease in the intercepts of curves.}
\end{figure*}

Second, the magnetic field changes the functional relationship between $S_{\rm AFM}^{\rm xx}$ and $U$ to suppress $S_{\rm AFM}^{\rm xx}$ in some ranges.
As shown in Fig.~\ref{FigSxU}(d), in contrast to the condition at $B = 0$ that $S_{\rm AFM}^{\rm xx}$ always has a positive correlation with $U$, with an increasing magnetic field, $S_{\rm AFM}^{\rm xx}$ as a function of $U$ has a decreasing area, and this area gradually expands. This is more clearly displayed in Fig.~\ref{FigSxU}(c), where $S_{\rm AFM}^{\rm xx}$ is plotted a function of $1/L$ under several values of $U$ at $B = 0.6$. In this panel, we observe that $U$ initially moves up (from 3.0 to 4.5) and subsequently moves down (from 4.5 to 6.0) the intercepts of the $S_{\rm AFM}^{\rm xx}$ curves. As the intercepts of the curves decrease, the antiferromagnetic phase is inhibited by the interaction rather than promoted.
We posit that the effect of interaction enhancing the spin polarization \textit{only} in a system with a magnetic field (as shown in Fig.~\ref{FigPB}(b))\cite{manolescu2011coulomb} will cause a system with $B \neq 0$ to gradually approach the fully spin-polarized state as U increases and is the reason for the novel phenomena regarding the change in the AFM phase.

Now, we further introduce disorder $\Delta$. We found that for the clean system, applying disorder suppresses the AFM phase. Under sufficiently strong $\Delta$, the AFM phase does not exist in the system regardless of how strong of a magnetic field is applied.
This phenomenon is visually demonstrated in Fig.~\ref{FigSxD}. In panel (a), in the process of $S_{\rm AFM}^{\rm xx}(B)$ increasing initially and subsequently decreasing, $|dS_{\rm AFM}^{\rm xx}/dB|$, which represents the degree of influence of the magnetic field on AFM, is greatly reduced or even eliminated by $\Delta$. When $\Delta$ reaches 1.2, there is almost no change in the $S_{\rm AFM}^{\rm xx}$ curve. The results with various lattice sizes are shown in panel (b), where $S_{\rm AFM}^{\rm xx}$ is a monotonically decreasing function of $\Delta$. This can be interpreted as the "screening effect" of the magnetic field, or the AFM phase, by the disorder $\Delta$. The change law of $S_{\rm AFM}^{\rm xx}$ with $\Delta$ can be regarded as basically the same under different lattice sizes.
The behavior with lattice sizes of $L = 6, 9, 12$ in the entire range of magnetic fields until full spin polarization is shown in Fig.~\ref{FigSxBD}. Since the relationship between $S_{\rm AFM}^{\rm xx}$ and $L$ determines the existence of the AFM phase, Fig.~\ref{FigSxBD} generally proves that the disorder suppresses the AFM phase. The increase in $\Delta$ directly leads to the pink curve, which reflects the intercepts of the $S_{\rm AFM}^{\rm xx}(1/L)$ curves, tending to zero. The pink curves of $\Delta=0.9$ and 1.2 in panels (c) and (d) are completely below the horizontal axis, it represents the complete disappearance of the AFM phase.
A similar effect of disorder was also confirmed in the Mott insulator, which was found to be caused by strong interaction\cite{PhysRevLett.120.116601,PhysRevB.64.184402,Otsuka_2000}, by showing the lack of an ordering wave vector in the randomness-dominated regime.

Although strong disorder always suppresses the AFM phase, its direct influence on the staggered transverse antiferromagnetic structure factor still needs to be examined in detail.
In Fig.~\ref{FigSxL}, we report $S_{\rm AFM}^{\rm xx}$ under several values of the disorder $\Delta$ as a function of $1/L$. For the metallic phase at $U = 3$ and $B = 0$ or the band insulating state at $U = 3$ and $B = 0.3$ (see the Appendix~\ref{app:DCcon}) in the clean system, $S_{\rm AFM}^{\rm xx}$ curves under various values of $\Delta$ are almost coincident, which shows that the effect of disorder can be considered non-existent regardless of whether there is a phase transition or not. As $B$ continues to increase to 0.6, disorder reduces the value to which the $S_{\rm AFM}^{\rm xx}(L)$ curve tends as $T \rightarrow 0$, thus the randomness it causes disrupts the magnetic order. Even at the stage in which $B$ reduces $S_{\rm AFM}^{\rm xx}$, this phenomenon of inhibiting the AFM phase is still effective, namely, the effect of disorder on $S_{\rm AFM}^{\rm xx}$ is not affected by the transport phases and only occurs when $B$ has a significant impact on the magnetic properties of the system.

\section{Summary}

Using DQMC simulations, we studied the magnetic phase transition of the disordered Hubbard model that is induced by a Zeeman field on a honeycomb lattice. For the magnetic order perpendicular to the lattice, applying a parallel field suppresses the possible AFM phase effectively, which can be induced by the Coulomb interaction. In the lattice plane, increasing $B$ first causes symmetry breaking and finally drives the system into a fully spin-polarized state, thereby leading to the trend of the $S_{\rm AFM}^{\rm xx}(B)$ curve initially rising and subsequently falling, which reflects the induction and inhibition of $B$ on the AFM phase.

In contrast to the weakening of the magnetic field effect by disorder, the interplay of $U$ and $B$ leads to more complex phenomena. The magnetic field completely changes the monotonic increase of $S_{\rm AFM}^{\rm xx}$ with $U$. Moreover, the peak value of the $S_{\rm AFM}^{\rm xx}$ curve, namely, $S_{max}$, and its corresponding magnetic field $B_{c}$ are both determined by the interaction, and the critical interaction for the AFM Mott insulating phase transition in the clean system also shows its particularity in this coupling effect: above and below this critical value $U_{c}$, $S_{max}$ has different functional relationships with the interaction. $U_{c}$ is approximately 4.5. Although this interaction is always greater than that in an actual material, the $B$-induced magnetic phase transition that occurs under small values of $U$ still provides a possibility for novel discoveries in experiments.

\noindent
\underline{\it Acknowledgments} ---
We thank Rubem Mondaini for many helpful discussions.
This work was supported by the NSFC (Nos. 11974049 and 11734002) and NSAF U1930402. The numerical simulations were performed at the HSCC of Beijing Normal University and on Tianhe-2JK in the Beijing Computational Science Research Center.

\appendix

\setcounter{equation}{0}
\setcounter{figure}{0}
\renewcommand{\theequation}{A\arabic{equation}}
\renewcommand{\thefigure}{A\arabic{figure}}
\renewcommand{\thesubsection}{A\arabic{subsection}}

\section{DC conductivity}
\label{app:DCcon}

To characterize the transport phase, we compute the $T$-dependent DC conductivity $\sigma_{\rm dc}$ via a proxy of the momentum $\bf q$ and imaginary time $\tau$-dependent current-current correlation function:
\begin{eqnarray}
\label{conduc}
\sigma_{\rm dc}(T)=\frac{\beta^{2}}{\pi}\Lambda_{xx}\left({\bf q}=0,\tau=\frac{\beta}{2}\right).
\end{eqnarray}
Here, $\Lambda_{xx}({\bf q},\tau)$ = $\langle \widehat{j_{x}}({\bf q},\tau)\widehat{j_{x}}(-{\bf q},0)\rangle$, and $\widehat{j_{x}}({\bf q},\tau)$ is the current operator in the $x$-direction. This form, which avoids the analytic continuation of the QMC data, has been shown to provide satisfactory results~\cite{Scalettar1999,PhysRevB.104.045138,Mondaini2012}. We implement the approach that is proposed in Ref. \cite{PhysRevB.54.R3756}, which is based on the following argument. The fluctuation-dissipation theorem yields the following:
\begin{eqnarray}
\label{conductivity}
\Lambda_{xx}(\textbf{q},\tau)=\frac{1}{\pi}\int d\omega \frac{e^{-\omega \tau}}{1-e^{-\beta\omega}}\text{Im}\Lambda_{xx}(\textbf{q},\omega),
\end{eqnarray}
where $\Lambda_{xx}$ is the current-current correlation function along the $x$-direction. While $\text{Im}\Lambda_{xx}(\textbf{q},\omega)$ could be computed by a numerical analytic continuation of $\Lambda_{xx}(\textbf{q},\tau)$ data that are obtained via the DQMC method, we instead assume here that $\text{Im}\Lambda_{xx}\sim\omega\sigma_{dc}$ below an energy scale $\omega < \omega^*$. If the temperature $T$ is sufficiently smaller than $\omega^*$, the above equation simplifies to
\begin{eqnarray}
\label{simpconduc}
\Lambda_{xx}\left(\textbf{q}=0,\tau=\frac{\beta}{2}\right)=\frac{\pi}{\beta^2}\sigma_{dc}
\end{eqnarray}
which is the form in Eq.~\eqref{conduc}.

This approach may not be valid for a Fermi liquid\cite{PhysRevB.54.R3756} when the characteristic energy scale is set by $\omega^* \sim N(0)T^2$, and the requirement $T<\omega^*$ will never be satisfied. However, in our system, the energy scale is set by the temperature-independent hopping-disorder strength $\omega^* \sim \Delta$; hence, Eq. (\ref{simpconduc}) is valid at low temperatures.

\begin{figure}[t]
\includegraphics[scale=0.5]{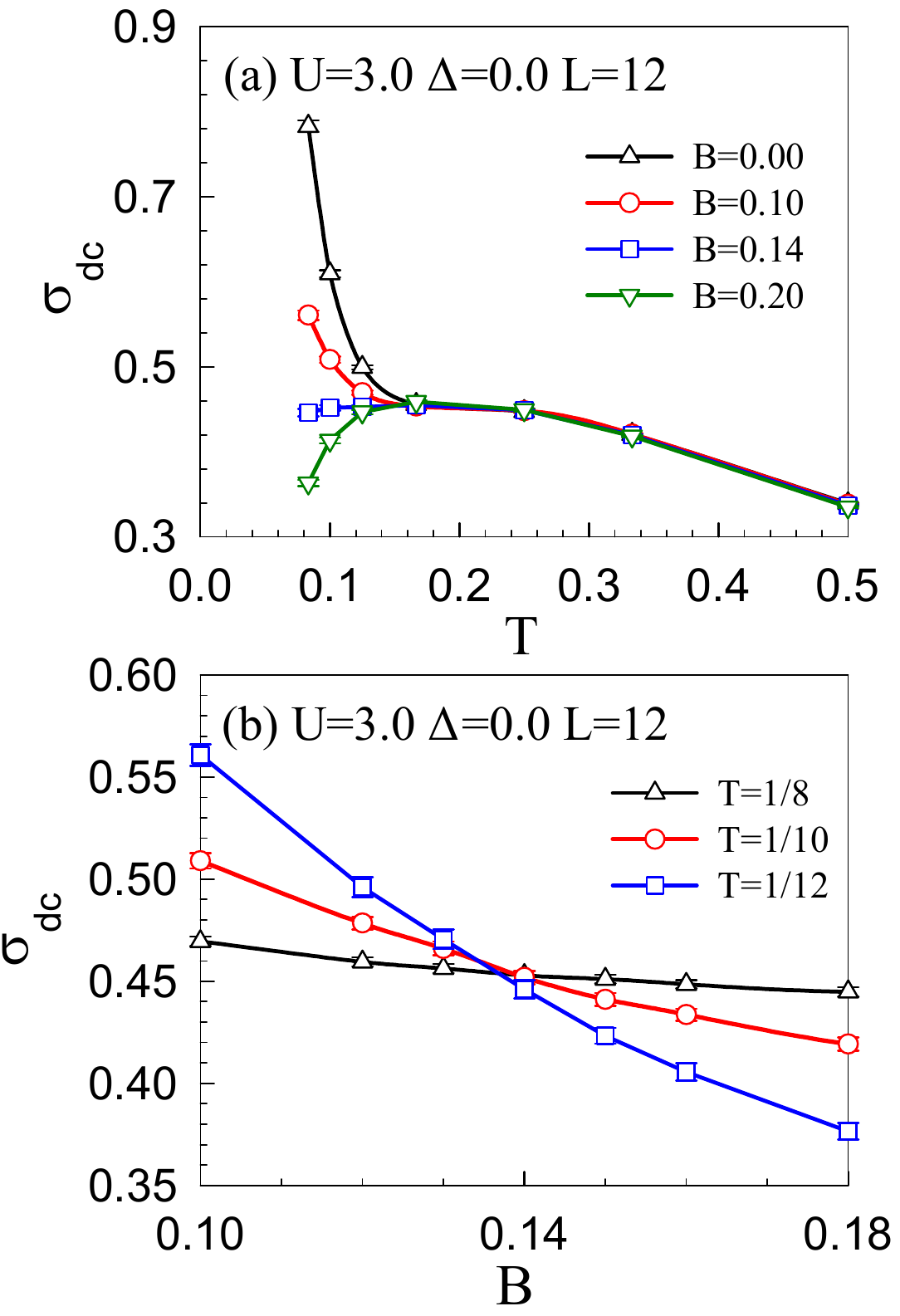}
\centering
\caption{\label{SMDcCon} The conductivity $\sigma_{\rm dc}$ is shown as a function of (a)temperature $T$ and (b) magnetic field $B$ at half-filling. Conductivity has different behaviors with the magnetic field greater/less than the critical value $B_{c}$ for the phase transition, and $B_{c}$ at $U = 3.0$ $\Delta = 0$ is about 0.14.
}
\end{figure}

As shown in Fig.~\ref{SMDcCon}, we use the low-temperature behavior of $\sigma_{\rm dc}$ to distinguish various transport properties. 
The system at $U=3.0$ and $\Delta=0.0$ changes from metal ($\sigma_{\rm dc}$ decreases with $T$) to insulator ($\sigma_{\rm dc}$ increases with $T$) as the magnetic field $B$ reaches 0.14, and it has been suggested that the phase transition is accompanied by the appearance of energy gap\cite{PhysRevB.104.045138}. However, although $B$ = 0.14 is sufficient to cause metal-insulator transition, it is too small to induce an AFM phase transition. As shown in Fig.~\ref{FigSxB}(b), when $B<$0.45, the AFM order does not appear, and the magnetic field had little effect on $S_{\rm AFM}^{\rm xx}$. Overall, $B=0.14$ is proven to introduce a band insulating phase into the system at $U=3$ and $\Delta=0$.

Different from the magnetic field, the critical interaction approximately equal to 4.5 can not only induce the AFM phase transition, but also opens the Mott gap\cite{PhysRevLett.120.116601,PhysRevB.72.085123}. Ref. \cite{1992U4.5,PhysRevB.101.245161} have shown that the interaction will induce a metal-Mott insulator transition in honeycomb lattice.

\section{Concerning the number of disorder realizations} \label{app:realz}
\begin{figure}[htbp]
\includegraphics[scale=0.5]{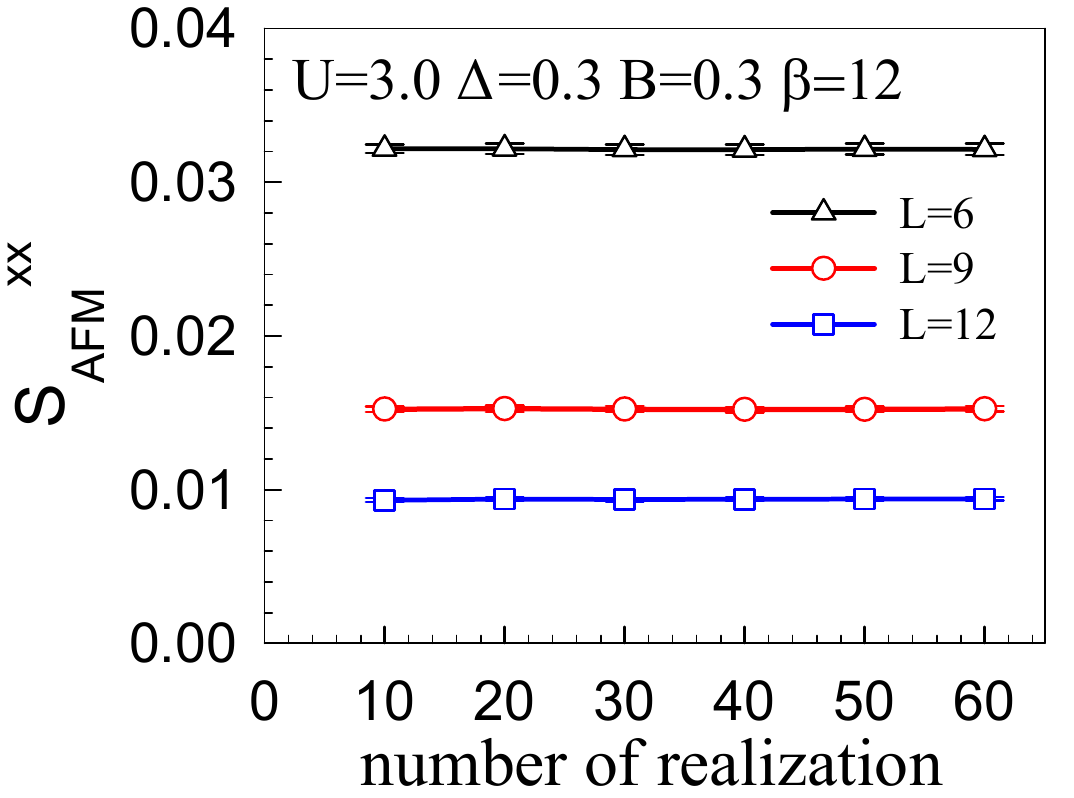}
\centering
\caption{\label{SMdel} $S_{\rm AFM}^{\rm xx}$ computed in the system with $\beta = $12, $U = $5, $\Delta=$0.5 and $B = $0.5.
For a specified value of $L$, the data that are obtained from an ensemble with an increasing number of disorder realizations are consistent within the statistical errors.
}
\end{figure}

In general, the required number of realizations in simulations with disorder must be determined empirically, which depends on a complex interplay among ``self-averaging'' on sufficiently large lattices, the disorder strength, and the location in the phase diagram. In Fig.~\ref{SMdel}, we show the results of $S_{\rm AFM}^{\rm xx}$ averaged over various numbers of random disorder realizations. For any specified lattice size $L$, the averaged $S_{\rm AFM}^{\rm xx}$ values are already consistent when $> 10$ realizations are used. This justifies the use of 20 realizations to obtain the results in the main text.

\bibliography{References}

\end{document}